\documentclass{ere04}

\usepackage{graphicx}    
\usepackage{psfrag}

\def\beq{\begin{equation}}
\def\eeq{\end{equation}}

\def\bea{\begin{eqnarray}}
\def\eea{\end{eqnarray}}
\def\ba{\begin{array}}                  
\def\ea{\end{array}}

\begin{document}

\title*{Spacelike slices from globally well-behaved simultaneity connections}
\author{E. Minguzzi\inst{1}}
\institute{Departamento de Matem\'aticas, Universidad de
Salamanca,  Plaza de la Merced 1-4, E-37008 Salamanca, Spain and
INFN, Piazza dei Caprettari 70, I-00186 Roma, Italy.
\texttt{minguzzi@usal.es} }
%
%
\maketitle

As shown by the development of Special Relativity the simultaneity
concept should be related to that of  reference frame. Poincar\'e
proposed to define the simultaneity of two events by means of
light signals following what is nowadays known as the Einstein
simultaneity convention. The need of a simultaneity definition is
present also in general relativity and in curved spacetimes in
order to provide the observers with a coordinate time.

It is recognized that the old Einstein simultaneity convention is
nothing but a connection on a suitable trivial bundle that defines
the reference frame. Unfortunately, it has a non vanishing
holonomy in curved and even in flat spacetimes a fact that makes
it almost useless. We point out the advantage of local
simultaneity conventions showing that they are represented by
local simultaneity connections. Among them there is one, uniquely
determined by the reference frame, which is particularly
well-behaved globally.

\section{Introduction}
At the beginning of the last century telegraphers were used to
synchronize distant clocks according to a procedure that took into
account the finite velocity of propagation of the telegraphic
signal \cite{galison03}. This procedure, later considered by
Poincar\'e \cite{poincare04a,poincare04b} for the propagation of
light signals, has become known as the Einstein synchronization
(simultaneity) convention. Consider two distant clocks $A$ and $B$
at rest in an inertial frame. The Einstein convention consists in
the following steps: An observer at $A$ (observer $A$ for short)
sends a light beam from $A$ to $B$ where it is reflected back to
$A$. Using its clock the observer at $A$ measures the round-trip
time $\Delta\tau$ of the light beam, and the instant of departure
$\tau_{Ai}$ while the observer at $B$ measures, with its own
clock, the time of reflection $\tau_{Br}$ that later is
communicated to $A$. $A$ then moves the hands of its own clock
forward of a quantity $\delta=\tau_{Br}-\Delta/2-\tau_{Ai}$ in
such way that if the entire process would have been done since the
beginning with the new setting $A$ would have found $\delta=0$.
The two clocks can be considered synchronized if this procedure,
no matter how many times repeated, gives always $\delta=0$. The
experience tells us that in fact, for clocks that have been
initially syntonized, if $\delta=0$ at one time then $\delta=0$ at
any later time. We recall that two clocks are syntonized if their
rate agree when moved in the same space point for the comparison.
From now on clocks will be assumed syntonized. Should we expect
that if $A$ and $B$ are synchronized and $B$ and $C$ are
synchronized then $A$ and $C$ are synchronized? That is, should we
expect that the Einstein synchronization convention in an inertial
frame is transitive? The answer is affirmative. It can be shown
\cite{weylG,minguzzi02d} that the transitiveness of the Einstein
convention follows only from the observable fact that the speed of
light over a closed path is a constant independent of the path
(here the light beam moves along the closed path by means of
suitable mirrors). It can therefore be safely applied to
synchronize all the clocks in space. With a lattice of clocks the
observers in the inertial frame are therefore able to construct a
coordinate time i.e. they are able to assign to each event its
coordinate time by simply looking at the nearest clock when the
event happens. The spacetime is thus foliated by simultaneity
slices of constant coordinate time, and the existence of such
slices is ultimately due to the transitiveness of the Einstein
simultaneity convention.

\section{The general relativistic picture}
The concepts and results of the previous section are,
unfortunately, only approximate since global inertial reference
systems do not exist. General relativity suggests that these
results could hold locally but not globally. Indeed it is easy to
show that the Einstein convention is not transitive, not even in
flat spacetime, if one considers rotating reference frames. First
of all one needs a concept for reference frame in general
relativity since the simultaneity concept should be related with
that of frame. Roughly speaking a reference frame is a collection
of objects (the points of the frame) moving together as a single
object. Mathematically the reference frame is therefore a
congruence of timelike curves \cite{hawking73,minguzzi03,manoff01}
on the curved spacetime $M$, that is, a reference frame is
determined by a normalized field $u(x)$.

The integral lines of this field are the worldlines of the points
at rest in the frame. If $S$ is the manifold of integral wordlines
we have (at least locally) a projection $\pi: M \to S$ that
associate to each event the worldline passing through it or, which
is the same, a point of space. Now, imagine applying the Einstein
convention on the neighborhood of an observer $A$ moving along the
congruence flow. Since in special relativity the Einstein
simultaneity convention determines as simultaneity planes those
perpendicular to the observer worldline the same should be true
locally in general relativity. Indeed, let $H_{u}(x)$ be the plane
orthogonal to $u(x)$ at $x$. This is the plane of simultaneity of
an observer moving with the flow, and thus having 4-velocity $u$,
in the sense that the events simultaneous to $x$ according to the
Einstein convention and with respect to the frame determined by
$u(x)$ lie near the exponential map at $x$ of $H_{u}(x)$.

Now, notice that a distribution of planes in a fibration $\pi: M
\to S$ determines a splitting of the tangent space $TM_x$ and
therefore defines a {\em generalized connection}. The conclusion
is that the Einstein simultaneity convention is in fact a
connection \cite{minguzzi03} in the sense of {\em generalized
gauge theories} \cite{michor91,modugno91}. Generalized gauge
theories have been studied extensively in the past since they have
a structure as rich as usual gauge theories without requiring a
principal bundle (that is the fibration is not necessarily
generated by the action of a group).

\begin{figure}[!ht]
\centering \psfrag{M}{$M$} \psfrag{H}{$M$}
\psfrag{I}{$\pi$}\psfrag{S}{$S$} \psfrag{G}{$\gamma$}
\psfrag{C}{$A$} \psfrag{L}{$S$} \psfrag{E}{$A_{1}$}
\psfrag{D}{$A_{2}$}
\includegraphics[width=7cm]{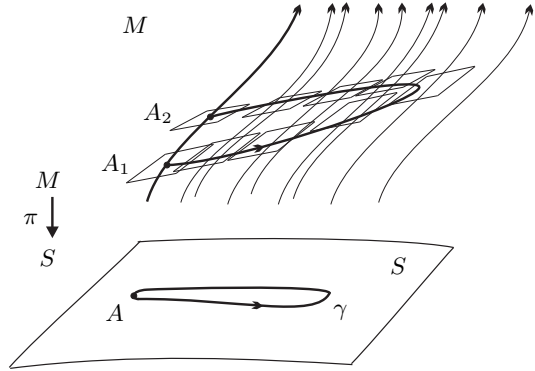}
\caption{The congruence of timelike curves, the base space $S$,
the horizontal planes, the horizontal lift and the holonomy.}
\label{fig1}
\end{figure}

It its interesting to look at the meaning of the horizontal lift
in this gauge theory of simultaneity \cite{minguzzi03}. Consider a
succession of observers at rest in the frame. Let $A$ be the first
observer and let observers B, C, D $\ldots$ lie around a closed
curve $\gamma$ on $S$. In other words let the observers be
disposed in circle and let them synchronize their clocks with that
of the observer at their own right-hand side. Starting from
observer $A$ the observers synchronize their clocks and thus
implicitly define a simultaneity convention which take as
simultaneous those events that correspond to the same clock
reading, no matter where placed along the curve. This
`simultaneous' events stay in a line which is the horizontal lift
of $\gamma$. The horizontal lift corresponds therefore to the
operation of pointwise synchronization along the base curve (see
figure \ref{fig1}).
Unfortunately, in presence of a non-vanishing holonomy the lifted
curve does not close. This means that the simultaneity considered
is not transitive along the closed curve and therefore it fails to
determine a spacetime foliation and hence a time coordinate. We
shall see, as a particular case of our study below, that the
Einstein simultaneity convention (connection) has a non vanishing
curvature (holonomy) whenever the vorticity vector
$w^{\eta}=\frac{1}{2}h^{\eta}_{\nu}\varepsilon^{\nu \beta \alpha
\gamma} u_{\beta} u_{\alpha;\gamma}$ differs from zero. Here
$h^{\eta}_{\nu}=\delta^{\eta}_{\nu}-u^{\eta}u_{\nu}$ is the
projector on the horizontal space $H_{u}$ (we use the timelike
convention $\eta_{00}=1$). Now, this vector field differs from
zero even in flat spacetime if a rotating congruence is
considered. The impossibility of determining a global simultaneity
definition using the Einstein convention in a rotating frame is
well known even experimentally. Indeed, the non-vanishing holonomy
implies that two beams of light moving over a closed curve
$\gamma$ in opposite directions close the curve in different
times. This is the celebrated Sagnac effect
\cite{sagnac13,ashtekar75}. These difficulties involving the
Einstein simultaneity convention and the practical need of a
coordinate time over rotating frames like the earth  naturally
arise the question as to whether some better simultaneity
convention could be found. The next section presents some results
obtained in \cite{minguzzi04}. We refer the reader to that work
for the detailed proofs.

\section{Alternative local simultaneity connections}
It is important to distinguish between local and non-local
simultaneity conventions. Non-local simultaneity conventions are
those that in order to construct a global spacetime foliation make
use of global information. This could be information on the global
shape and metric of the spacetime manifold (say the Schwarzschild
metric) or on the motion of distant objects (say the satellites
used in the GPS). Local simultaneity conventions are those
conventions that hopefully determine a global spacetime foliation
starting from local information only. They apply as rules between
neighboring observers exactly as the Einstein convention does. The
single observer at rest in the frame does not need global
information on the frame or on the spacetime metric structure.
These are the most conceptually easier simultaneity conventions,
but despite of this, conventions of this kind were almost
overlooked in previous literature with the notable exception of
the Einstein convention. Mathematically a local simultaneity
convention is, analogously to the Einstein case, a connection over
the frame bundle, that is a splitting of the tangent space in
vertical and horizontal. This can be determined in a number of
ways. Here we shall identify the connection with a 1-form $\omega$
normalized so that $\omega(u)=1$ and such that $\omega^{\mu}$ is a
timelike vector (Llosa and Soler would call our connection
$\omega$ a normalized {\em time scale} \cite{llosa04}). This last
condition is imposed since the horizontal space $H_{\omega}(x)$ at
$x$ is identified with the ker of $\omega$ at $x$, and the
previous condition assures that it is spacelike, a minimal
requirement for a simultaneity plane. This defines a general
simultaneity connection. However, we  are interested in {\em
local} simultaneity connections. This requirement imposes some
more conditions on the shape of $\omega$ and its meaning should be
briefly discussed first. We have already said that a local
convention should use only local information that the generic
observer at rest should find with experiments in its local
comoving laboratory. This information may consist in data on the
motion of the congruence as the 4-velocity $u^{\mu}$, the
vorticity vector $w^{\mu}$, the acceleration
$a^{\mu}=u_{\mu;\alpha}u^{\alpha}$, the shear tensor, the
expansion and in data on the  spacetime metric structure such as
the Riemann tensor. A local simultaneity connection (convention)
is therefore a connection $\omega$ that can be contructed from
local tensors. Exotic tensors can in principle enter the
contruction, however, this is unlikely since they should have a
clear operational meaning. That is, it should be clear what the
observer should do in order to measure their components in a
suitable base. It is convenient to introduce the vector product
between the vorticity vector and the acceleration
$m_{\alpha}=\epsilon_{ \alpha \beta \gamma \delta} a^{\beta}
u^{\gamma} w^{\delta}$ and limit for the moment our analysis only
to those spacetime regions where $m_{\alpha}\ne 0$. We also define
$a^2=-a^{\mu} a_{\mu}$, $w^2=-w^{\mu}w_{\mu}$ and
$m^{2}=-m^{\mu}m_{\mu}=a^{2}w^{2}\sin^{2}\theta$ where $\theta$ is
the angle between the vorticity vector and the acceleration for an
observer moving at speed $u$. Since $u^{\mu}$, $a^{\mu}$,
$w^{\mu}$ and $m^{\mu}$ are linearly independent any local
simultaneity convention takes the form
\begin{equation}\label{generic}
\omega_{\alpha}=u_{\alpha}+\psi^{m}(x) m_{\alpha}+\psi^{a}(x)
a_{\alpha}+ \psi^{w}(x) w_{\alpha} ,
\end{equation}
for suitable functions $\psi^{m},\psi^{a},\psi^{w}$. From the
definition of local simultaneity convention it follows moreover
that $\psi^{m},\psi^{a},\psi^{w}$, depend on the acceleration $a$
the vorticity $w$, the angle $\theta$, and possibly on other
scalars (note that in a stationary frame the possibilities are
reduced since the shear and the expansion vanish). Note that if
the $\psi$ functions are small the simultaneity connection may be
considered as a perturbation of Einstein's for which
$\omega_{\alpha}=u_{\alpha}$.

Let us come to the curvature. Here, for short, we identify it with
the vector (for the relations between different definitions see
\cite{minguzzi04})
\begin{equation}
v^{\eta}=\frac{1}{2}h^{\eta}_{\nu}\varepsilon^{\nu \beta \alpha
\gamma} \omega_{\beta} \omega_{\alpha;\gamma}
\end{equation}
It is a kind of generalized  vorticity vector for the present
non-time orthogonal connection. Taking into account the Frobenius
condition of integrability, $\omega \wedge d \omega=0$, it is not
difficult to show that the equation $v^{\mu}=0$ implies that the
distribution of planes $H_{\omega}$ in the ker of $\omega$ is
integrable.

\begin{figure}[!ht]
\centering \psfrag{U}{$ \ u$} \psfrag{H}{$H_{u}$}
\psfrag{K}{$H_{\omega}$}
\includegraphics[width=5cm]{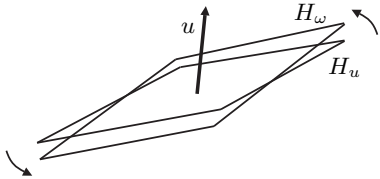}
\caption{Can the horizontal planes of the Einstein simultaneity
convention be slightly tilted according to a local rule so as to
obtain a new integrable distribution of planes?} \label{fig2}
\end{figure}

Since in the Einstein case the curvature coincides with the
vorticity vector we  might look for new connections that reduce to
Einstein's if $w^{\mu}=0$. Hence the additional term $\psi^{m}(x)
m_{\alpha}+\psi^{a}(x) a_{\alpha}+ \psi^{w}(x) w_{\alpha}$ in the
expression for $\omega$ should vanish whenever the vorticity
vanish. The idea is that of looking for some functions $\psi$ that
tilting suitably the simultaneity planes make them integrable (see
figure \ref{fig2}).
However, notice that since the functions $\psi$ depend on $x$ only
indirectly through the dependence on some measurable scalars we
are actually looking for a local rule of tilting. In practice, in
order to apply the simultaneity convention $\omega$, the observers
at rest in the frame apply the same procedure using light rays of
the Einstein convention. This time, however, they slightly modify
the coefficient $\delta$. Rewrite the connection $\omega_{\mu}$ as
$\omega_{\mu}=u+\phi n_{\mu}$ where $n$ is a spacelike normalized
vector. Since the functions $\psi^{m}$, $\psi^{a}$ and $\psi^{w}$
are measurable so are $\phi(x)$ and $n(x)$. Let us return to two
neighboring observers A and B and consider a light beam sent from
A to B and then reflected back to A. Let $\alpha$ be the angle
between the direction $AB$ and $n$. With the previous notation, in
order to apply the convention $\omega$ the observer $A$ should
move forward the hands of its clock of a quantity (for simplicity
we are considering here the case in which there is no redshift
between A and B; this formula can, however, be generalized to
include the
redshift)$\delta_{\omega}=\tau_{Br}-\frac{\Delta}{2}-\tau_{Ai}+\phi
\Delta/2 \cos \theta =\delta_{u}+\frac{1}{2}\phi \Delta \cos
\theta$ with $\delta_{u} \equiv \delta$.

Let us now try to determine the most convenient local simultaneity
connection. We make some simplifying assumptions
\begin{itemize}
\item[(a)] The frame is generated by a Killing vector field $k$.
\item[(b)] The functions $\psi^{m}$, $\psi^{w}$ and $\psi^{a}$
are constructed from the observable quantities $a$, $w$ and
$\theta$ (or equivalently  $a$, $w$ and $m$ with $m=a w \sin
\theta$).
\item[(c)] The curvature $v^{\eta}$ of $\omega_{\mu}$ is proportional
to the Riemann tensor (through contraction with a suitable
tensor).
\end{itemize}
The first two conditions are natural simplifications that allow us
to tackle the problem while keeping the calculations at a
reasonable size. The last one is imposed since the requirement
$v^{\mu}=0$ would be too restrictive and no simultaneity
connection satisfying that requirement would be eventually found
(note that the condition $v^{\mu}=0$ implies condition (c), thus
if a connection that satisfies $v^{\mu}=0$ exists, it should be
found between those that we selected imposing (c)). With our
condition (c), at least in the weak field limit, the distribution
of horizontal planes becomes integrable providing a useful
definition of simultaneity.

The following theorem holds \\
\begin{theorem}
In a stationary spacetime let $k$ be a timelike Killing vector
field and set $u=k/\sqrt{k \cdot k}$. Let $U$ be the open set
$U=\{x: m(x)
> 0 \ \textrm{and} \ a(x) \ne w(x) \}$. Consider in $U$ the connection
\begin{equation} \label{conn}
\omega_{\alpha}=u_{\alpha}+\psi^{m}(x) m_{\alpha}+\psi^{a}(x)
a_{\alpha}+ \psi^{w}(x) w_{\alpha}.
\end{equation}
Let $\psi^{m},\psi^{a},\psi^{w}$, be $C^{1}$ functions dependent
only on $a$, $w$ and $\theta$.  Then, regardless of the stationary
spacetime considered, the connection is timelike in $U$ (and hence
it is a simultaneity connection in $U$) and has a curvature
proportional to the Riemann tensor in $U$  only if
\begin{equation} \label{magic}
\psi^{m}=\frac{a^{2}+w^{2}-\sqrt{(a^{2}+w^{2})^{2}-4m^{2}}}{2m^{2}}. \\
\end{equation}
\end{theorem}

\section{Conclusions}
Thanks to the previous theorem the simultaneity connection
\begin{equation} \label{cbar}
\bar{\omega}_{\alpha}=u_{\alpha}+
\frac{a^{2}+w^{2}-\sqrt{(a^{2}+w^{2})^{2}-4m^{2}}}{2m^{2}}\,
m_{\alpha},
\end{equation}
that we call $\bar{C}$-simultaneity, is particularly well behaved
in those spacetime regions where  the Riemann tensor is
sufficiently weak. It can be shown \cite{minguzzi04} that it can
be extended by continuity to the set $C=A-B$ where, $A=\{x:
a^{2}+w^{2}>0\}$, $B=\{x: a=w\ne 0 \ \textrm{and} \
\theta=\pi/2\}$, by defining, $\bar{\omega}_{\alpha}=u_{\alpha}$,
in those points where $m=0$.

The $\bar{C}$-simultaneity follows almost uniquely from the
geometrical requirements discussed above. While the observers have
the freedom to choose a local simultaneity convention, the
requirement of being almost integrable in the weak field limit
imposes strong constraints on its actual expression. The geometry
tells the observers what simultaneity convention is better to use
in practice and remarkably the simultaneity convention that turns
out is not Einstein's but rather $\bar{C}$-simultaneity.
\begin{figure}[!t]
\centering
\includegraphics[width=7cm]{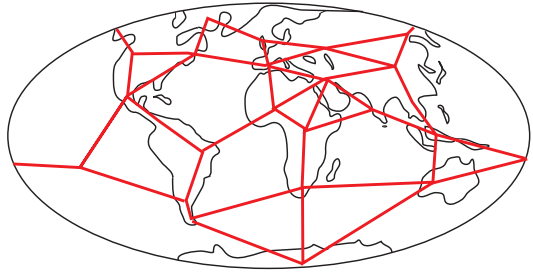}
\caption{A network of observers over the earth surface.}
\label{fig3}
\end{figure}
This convention is integrable in Minkowski spacetime contrary to
the Einstein convention which is not integrable, for instance, in
the case of the rotating platform. This integrability can also be
verified studying the more complicated Killing vector field that
gives rise to the pseudocylindric coordinates  of Letaw and
Pfautsch \cite{letaw81,letaw82}.

We believe that our approach shares a number of features that
makes it preferable over other approaches to simultaneity. Indeed
it is coordinate independent, it has a clear operational meaning,
and the locality property makes it independent of the global
information that one may or may not have (e.g. it does not depend
on the model of the earth geoid). Finally, it does not require
particular spacetime symmetries apart from that of stationarity
(e.g. the spherical symmetry is not required). Our findings seems
to be useful for all those synchronization approaches that try to
adapt a global coordinate time to a spacetime network of observers
(e.g. computers connected over the earth surface) without
privileging a particular vertex in the network and without making
use of elements outside the network (e.g. the satellites in the
GPS).

\section*{Acknowledgements}
I thank the department of mathematics of  Salamanca for kind hospitality. This work has been supported by INFN, grant $\textrm{n}^{\circ}$ 9503/02.\\

%
%
%
%

\begin{thebibliography}{10}

\bibitem{ashtekar75}
A.~Ashtekar and A.~Magnon.
\newblock The {S}agnac effect in general relativity.
\newblock {\em J. Math. Phys.}, 16:341--344, 1975.

\bibitem{galison03}
P.~Galison.
\newblock {\em Einstein's Clocks, {P}oincar\'e's Maps: {E}mpires of Time}.
\newblock Norton and Company, New {Y}ork, 2003.

\bibitem{hawking73}
S.~W. Hawking and G.~F.~R. Ellis.
\newblock {\em The Large Scale Structure of Space-Time}.
\newblock Cambridge {U}niversity {P}ress, Cambridge, 1973.

\bibitem{letaw81}
J.~R. Letaw.
\newblock Stationary world lines and the vacuum excitation of noninertial
  detectors.
\newblock {\em Phys. Rev. D}, 23:1709--1714, 1981.

\bibitem{letaw82}
J.~R. Letaw and J.~D. Pfautsch.
\newblock The stationary coordinate systems in flat spacetime.
\newblock {\em J. Math. Phys.}, 23:425--431, 1982.

\bibitem{llosa04}
J.~Llosa and D.~Soler.
\newblock Reference frames and rigid motions in relativity.
\newblock {\em Class. Quantum Grav.}, 21:3067--3094, 2004.

\bibitem{manoff01}
S.~Manoff.
\newblock Frames of reference in spaces with affine connections and metrics.
\newblock {\em Class. Quantum Grav.}, 18:1111--1125, 2001.

\bibitem{michor91}
P.~W. Michor.
\newblock {\em Gauge theory for fiber bundles}, volume~19 of {\em Monographs
  and Textbooks in Physical Sciences}.
\newblock Bibliopolis, Napoli, 1991.

\bibitem{minguzzi03}
E.~Minguzzi.
\newblock Simultaneity and generalized connections in general relativity.
\newblock {\em Class. Quantum Grav.}, 20:2443--2456, 2003.

\bibitem{minguzzi04}
E.~Minguzzi.
\newblock A globally well-behaved simultaneity connection for stationary frames
  in the weak field limit.
\newblock {\em Class. Quantum Grav.}, 21:4123--4146, 2004.

\bibitem{minguzzi02d}
E.~Minguzzi and A.~Macdonald.
\newblock Universal one-way light speed from a universal light speed over
  closed paths.
\newblock {\em Found. Phys. Lett.}, 16:587--598, 2003.

\bibitem{modugno91}
M.~Modugno.
\newblock Torsion and {R}icci tensor for non-linear connections.
\newblock {\em Differential Geometry and its Application}, 1:177--192, 1991.

\bibitem{poincare04a}
H.~Poincar{\'e}.
\newblock {\em Bull. des Sci. Math.}, 28:302, 1904.

\bibitem{poincare04b}
H.~Poincar{\'e}.
\newblock {\em La Revue des I{d\'e}es}, 1:801, 1904.

\bibitem{sagnac13}
M.~G. Sagnac.
\newblock {\em C. R. Acad. Sci.}, 157:708, 1913.
\newblock ibid. 1410.

\bibitem{weylG}
Hermann Weyl.
\newblock {\em Raum Zeit Materie}.
\newblock {Springer-Verlag}, New York, 1988.
\newblock Seventh edition based on the fifth {G}erman edition (1923).

\end{thebibliography}
\end{document}